\documentclass[aps, prb, reprint, superscriptaddress,
amssymb, amsmath]{revtex4-2}

\usepackage[utf8]{inputenc}
\usepackage[english]{babel}
\usepackage{graphicx}
\usepackage{hyperref}
\usepackage{bm}
\usepackage{color}

\DeclareMathOperator{\Tr}{\mathrm{Tr}}
\renewcommand{\Re}{\operatorname{Re}}
\renewcommand{\Im}{\operatorname{Im}}

\makeatletter
\newsavebox{\@brx}
\newcommand{\llangle}[1][]{\savebox{\@brx}{\(\m@th{#1\langle}\)}%
	\mathopen{\copy\@brx\kern-0.5\wd\@brx\usebox{\@brx}}}
\newcommand{\rrangle}[1][]{\savebox{\@brx}{\(\m@th{#1\rangle}\)}%
	\mathclose{\copy\@brx\kern-0.5\wd\@brx\usebox{\@brx}}}
\makeatother

\begin{document}
\title{Level attraction and exceptional points in a resonant spin-orbit
torque system}

\author{Igor \surname{Proskurin}}
\email{Igor.Proskurin@umanitoba.ca}
\affiliation{Department of Physics and Astronomy, University of Manitoba,
Winnipeg, MB R3T~2N2, Canada}
\affiliation{Institute of Natural Sciences and Mathematics, Ural Federal
University, Ekaterinburg 620002, Russia}

\author{Robert L. \surname{Stamps}}
\email{Robert.Stamps@umanitoba.ca}
\affiliation{Department of Physics and Astronomy, University of Manitoba, 
Winnipeg, MB R3T~2N2, Canada}

\begin{abstract}
Level attraction can appear in driven systems where instead of repulsion two
modes coalesce in a region separated by two exceptional points.  This behavior
was proposed for optomechanical and optomagnonic systems, and recently observed
for dissipative cavity magnon-polaritons.  We demonstrate that such a regime
exists in a spin-orbit torque system where a magnetic oscillator is resonantly
coupled to an electron reservoir.  An instability mechanism necessary for mode
attraction can be provided by applying an electric field.  The field excites
interband transitions between spin-orbit split bands leading to an instability
of the magnetic oscillator.  Two exceptional points then appear in the
oscillator energy spectrum and the region of instability.  We discuss conditions
under which this can occur and estimate the electric field strength necessary
for reaching the attraction region for a spin-orbit torque oscillator with
Rashba coupling. A proposal for experimental detection is made using magnetic
susceptibility measurements.
\end{abstract}

\maketitle

\section{Introduction}
In noncentrosymmetric systems, the spin-orbit interaction provides mechanisms
for spin angular momentum transfer and generation of spin-orbit torques
\cite{Manchon2019}.  Microscopically, this can be traced back to the spin Hall
\cite{Sinova2015} and inverse spin galvanic effects \cite{Dyakonov1971,
Edelstein1990}, where an electric field, can generate either a spin current or a
spin polarization. In this paper we show how resonant spin-orbit torque devices
can reveal aspects common to driven interacting systems such as mode attraction
and exceptional points.

Spin-orbit torques play an important role in spintronics \cite{Manchon2008,
Chernyshov2009, Manchon2009, Matos2009, Miron2010, Gambardella2011, Wang2012,
Hals2013, Hals2015, Freimuth2015, Sokolewicz2019, Haku2020, Hibino2020,
Filianina2020, Saha2020}, with numerous applications including electric control
and magnetization switching \cite{Liu2012, Wadley2016, Shi2020}, coherent
excitation and amplification of spin waves \cite{Gladii2016, Collet2016,
Demidov2017, Demidov2020}, current-induced collective dynamics of topological
spin textures \cite{Emori2013, Haazen2013, Martin2020, Sanchez2020, Hanke2020},
memory and logic devices \cite{Bhowmik2014, Olejnik2017, Luo2020}, and
neuromorphic computing \cite{Torrejon2017}.  Spin-orbit torque oscillators are
of particular interest for many applications \cite{Demidov2012, Duan2014,
Cheng2016, Evelt2018, Haidar2019}, including damping and antidamping torques
\cite{Demidov2011} and other nontrivial features in high frequency response
\cite{Berger2018, Berger2018a}.

Level attraction can occur at mode energy level crossings of an open system. It
requires an instability of one of the modes, and in contrast to mode
hybridization where two energy levels repel each other, energy levels coalesce
for attraction, which in a certain way is reminiscent of mode synchronization
\cite{Bernier2018}. The region of mode coalescence is bounded by two exceptional
points where the eigenvectors become parallel in Hilbert space \cite{Heiss2004,
Heiss2012}. These points are singularities and have been observed in microwave
cavity applications.  Their nontrivial topology may have utility in detecting
devices \cite{Chen2017, Hodaei2017, Zhong2019}.

Additionally, mode attraction has been studied in optomechanical circuits
\cite{Bernier2018}, cold atoms with negative mass instability \cite{Kohler2018},
and coupled spin-photon systems \cite{Grigoryan2018, Proskurin2019, Yu2019,
Grigoryan2019, Karg2020, Tserkovnyak2020}, and has been experimentally
demonstrated for dissipative cavity magnon-polaritons \cite{Harder2018,
Yang2019, Wang2019, Rao2019}.

In this paper we show that a driven spin-orbit torque oscillator has a strong
interaction regime where ferromagnetic resonance can be locked to interband
electron transitions in a material with spin-orbit coupling.  The exceptional
points that appear can be electrically controlled.  The nontrivial topological
structure of the parameter space near the exceptional points \cite{Heiss2012}
affects mode selection and is sensitive to damping and coupling strength.

The instability mechanism needed to reach the attractive regime is provided by
spin accumulations created by an electric field applied to the electron system
in a manner similar to inverse spin galvanic spin-orbit torques
\cite{Manchon2008, Chernyshov2009}.  We show that in a driven system spin
accumulations modify an effective coupling between the resonances in magnetic
and electron subsystems. As a result, the system near the energy level crossing
can be described by a non-Hermitian Hamiltonian with a complex interaction
constant.

Mode attraction can be experimentally detected by measuring magnetic
susceptibility of a spin-orbit torque oscillator in the energy range where the
ferromagnetic resonance of the magnetic oscillator is close to that of the
electron band gap of the material with spin-orbit interaction.  For a spin-orbit
torque oscillator with Rashba coupling, we show that if the electric field is
above some critical value, behavior of the susceptibility changes drastically in
the region where magnetic and electron resonances strongly interact. Similar
response has been reported for spin-orbit torque multi-layers in microwave
waveguides \cite{Berger2018}.  We also demonstrate that it is possible to
navigate around an exceptional point and explore its topological structure by
sweeping strength and direction of the electric field. We estimate the critical
field and discuss symmetry conditions required to realize this behavior in a
Rashba system.

The paper is organized as follows.  In Sec.~\ref{sec:int}, we briefly review
mode repulsion and attraction.  A model for the spin-orbit torque oscillator is
introduced in Sec.~\ref{sec:mod}.  In Sec.~\ref{sec:tls}, we discuss general
properties of the magnetic susceptibility in an open spin-orbit torque system.
These results are used in Sec.~\ref{sec:lat} to explain the level attraction
regime, and are applied to the Rashba spin-orbit torque oscillator in
Sec.~\ref{sec:res}, where we discuss possibilities for experimental detection.
Concluding remarks are given in Section~\ref{sec:con}.

\section{Level attraction and exceptional points at a level crossing of an open system}
\label{sec:int}
We begin with a description of a physical picture at the mode energy level
crossing of an open system.  Suppose we have a mode described by the ladder
operators $a$ and $a^{\dag}$ and the frequency $\Omega$.  Its interaction with
another mode near the energy level crossing can be described by a linear
Hamiltonian
\begin{equation}\label{eq:osc}
	H_{\ell} = \hbar\Omega a^{\dag}a + \hbar\Delta c^{\dag} c + 
	\hbar g(c^{\dag} a + a^{\dag}c),
\end{equation}
where $c$ and $c^{\dag}$ are the ladder operators for the second mode, $\Delta$
is its frequency, and $g$ describes interaction between two modes.  The spectrum
of this interacting system is calculated as
\begin{equation}\label{eq:eif}
	\omega^{(\pm)} = \frac{\tilde{\Omega} + \tilde{\Delta}}{2} \pm \sqrt{\left( \frac{\tilde{\Omega} - \tilde{\Delta}}{2} \right)^{2} + g^{2}},
\end{equation}
where we included phenomenological mode damping parameters $\Gamma$ and $\gamma$
into the complex frequencies $\tilde{\Omega} = \Omega - i\Gamma$ and
$\tilde{\Delta} = \Delta - i\gamma$.

\begin{figure}[t]
	\centerline{\includegraphics[width=0.485\textwidth]{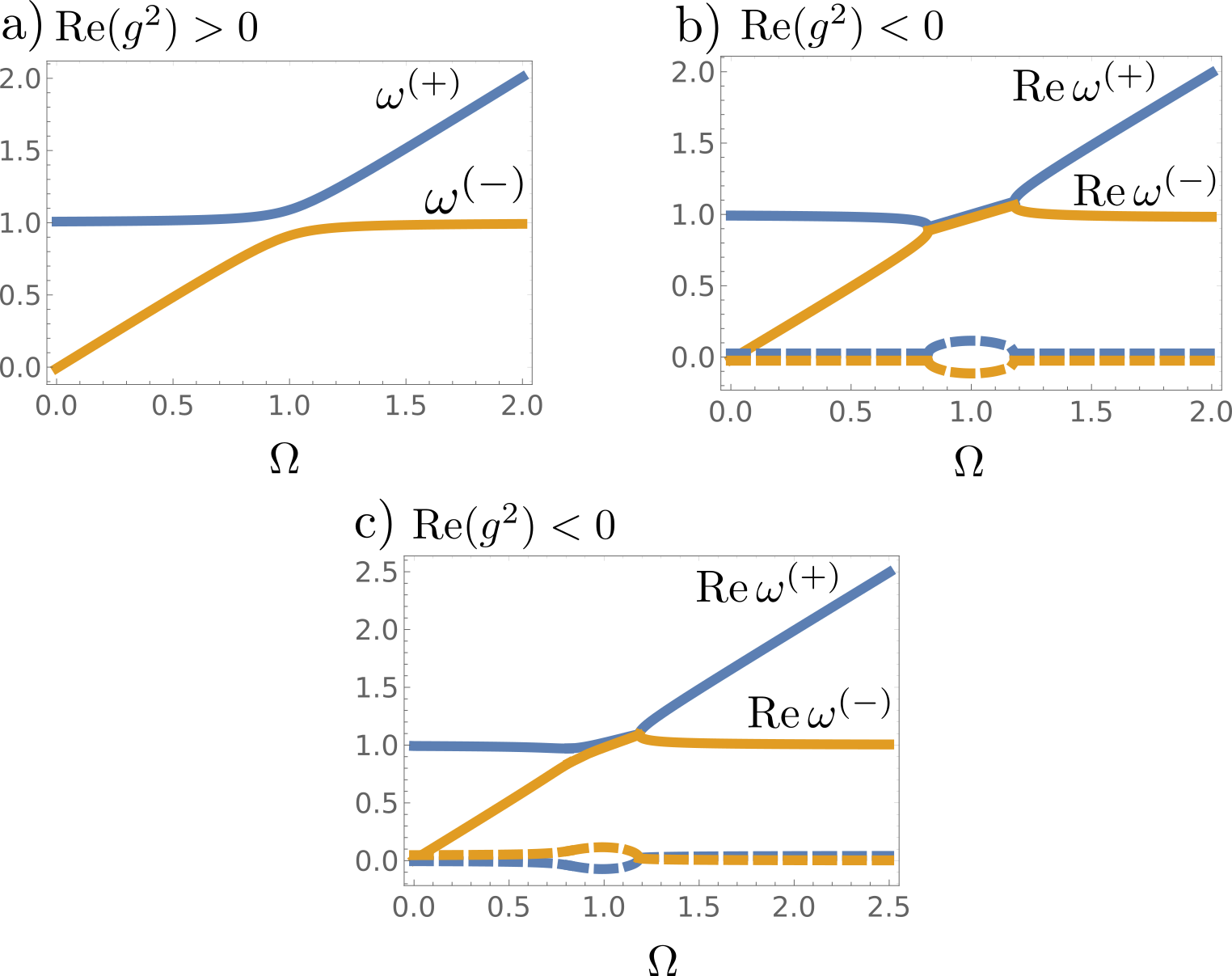}}
	\caption{Level repulsion and attraction. (a) Typical hybridization picture
	between energy levels $\Delta = 1$ and  $\Omega$ in the case of $g^{2} > 0$. (b)
	Level attraction picture with $\Re (g^{2}) < 0$ and $\Im (g^{2}) =0$ when
	$\Gamma = \gamma$. The energy modes $\omega^{(\pm)}$ coalesce in the region
	$(\Omega - \Delta) < 4 g^{2}$ bounded by two exceptional points.  (c) 
	Mode attraction is broken when $\Im (g^{2}) \ne 0$. One exceptional point,
	which satisfies Eqs.~(\ref{eq:ex}), is realized. Dashed lines show the
	imaginary parts of $\omega^{(\pm)}$.}
	\label{fig:1}
\end{figure}

For an isolated system in thermodynamic equilibrium, $g$ is a real parameter,
which ensures that $H_{\ell}$ is a Hermitian operator, so that Eq.~(\ref{eq:eif})
describes usual hybridization between two modes shown in Fig.~\ref{fig:1}~(a)
with a gap proportional to $g^{2}$.  As we show later on the example of a
spin-orbit torque oscillator, level crossing of an open system can be also
effectively described by Eq.~(\ref{eq:osc}).  However, the interaction parameter
$g$ in this case may become complex as result of interaction with an
environment.

Let us consider a situation where $g$ is complex.  While $\Re(g^{2})$ remains
positive, the interation picture at the level crossing remains qualitatively the
same as in Fig.~\ref{fig:1}~(a).  This situation however changes drastically if
$\Re(g^{2})$ becomes negative.  It indicates an instability developing in the
coupled system.

A detailed picture depends also on the imaginary part of $g^{2}$. If we can
neglect $\Im(g^{2})$, in the region $|\Omega - \Delta| < 4 |g^{2}| $ the system
enters a regime where $\omega^{(+)}$ and $\omega^{(-)}$ coalesce (see
Fig.~\ref{fig:1}~b).  Two exceptional points, where the Hamiltonian in
Eq.~(\ref{eq:osc}) is not diagonalizable, mark the boundaries of this region. To
realize this scenario, it is important to tune the dissipation channels so that
$\Gamma \approx \gamma$ \cite{Bernier2018}.

The imaginary part of $g^{2}$ modifies this behavior, and attraction may not be
fully realized (see Fig.~\ref{fig:1}~c).  Nevertheless, exceptional points still
exist, and are determined from the equation $(\tilde{\Omega} -
\tilde{\Delta})^{2} = -4 g^{2}$, which is satisfied for
\begin{subequations}
	\label{eq:ex}
	\begin{align}
		\label{eq:ex1}
		\frac{1}{2}(\Omega - \Delta)^{2} &= |g^{2}| + |\Re(g^{2})|, \\ 
		\label{eq:ex2}
		\frac{1}{2}(\Gamma - \gamma)^{2} &= |g^{2}| - |\Re(g^{2})|.
	\end{align}
\end{subequations}
The second equation here can be interpreted as a balance of energy dissipation in the system.

In the next sections, we show how this scenario can be realized for a spin-orbit
torque oscillator driven by an electric field.

\section{The model for a spin-orbit torque system}
\label{sec:mod}
Resonant properties of a spin-orbit torque system can be described by a model
\cite{Manchon2019, Manchon2008, Wang2012} where conducting electrons in energy
band $\varepsilon_{\bm{p}}$ interact with spin magnetic moment $\bm{S}$ via an
$s$-$d$ exchange field $\Delta$ in the presence of a spin-orbit interaction
$\bm{l}_{\bm{p}}$
\begin{equation}
H = \sum_{\bm{p}} \varepsilon_{\bm{p}} \psi^{\dag}_{\bm{p}} \psi_{\bm{p}} + \sum_{\bm{p}} \bm{l}_{\bm{p}}\cdot \psi^{\dag}_{\bm{p}} \bm{\sigma} \psi_{\bm{p}} - \Delta \bm{S} \cdot  \sum_{\bm{p}}  \psi^{\dag}_{\bm{p}} \bm{\sigma} \psi_{\bm{p}},
\label{eq:H0}
\end{equation}
where $\sigma_{i}$ ($i=x,y,z$) are the Pauli matrices, and $\psi_{\bm{p}} =
(\psi_{\bm{p}\uparrow}, \psi_{\bm{p}\downarrow})^{T}$, $\psi^{\dag}_{\bm{p}} =
(\psi^{\dag}_{\bm{p}\uparrow}, \psi^{\dag}_{\bm{p}\downarrow})$, are the
electron creation and annihilation operators with momentum $\bm{p}$ and spin
$\sigma = (\uparrow, \downarrow)$.

We decompose $\bm{S}$ into a static part $\bm{S}_{0}$ and small dynamic
perturbation $\delta\bm{S}$.  By choosing $\bm{S}_{0}$ along the $z$ direction
($S_{0} = 1$ hereafter), the static part of the Hamiltonian is written in the
following form
\begin{equation}
H_{0} = \sum_{\bm{p}} \psi^{\dag}_{\bm{p}} 
\begin{pmatrix}
\varepsilon_{\bm{p}} - \Delta' & l_{\bm{p}}^{(-)} \\
l_{\bm{p}}^{(+)} & \varepsilon_{\bm{p}} + \Delta'
\end{pmatrix}
\psi_{\bm{p}},
\label{eq:H1}
\end{equation}
where $\Delta' = \Delta - l^{z}_{\bm{p}}$ and $l_{\bm{p}}^{(\pm)} =
l_{\bm{p}}^{x} \pm il_{\bm{p}}^{y}$.  The energy spectrum can be found by
transforming to the basis where $H_{0}$ is diagonal, $\psi_{\bm{p}} = U_{\bm{p}}
c_{\bm{p}}$, with the unitary transformation
\begin{equation}
	\label{eq:Up}
	U_{\bm{p}} =
	\begin{pmatrix}
		\dfrac{l_{\bm{p}}^{(-)}}{\Delta_{\bm{p}}} & -\dfrac{l_{\bm{p}}^{(-)}}{\Delta_{\bm{p}}} \\
		\dfrac{\Delta_{\bm{p}} + \Delta'}{\sqrt{N^{(+)}}}                                                 & \dfrac{\Delta_{\bm{p}} - \Delta'}{\sqrt{N^{(-)}}} 
	\end{pmatrix},
\end{equation}
where $N^{(\pm)} = 2\Delta_{\bm{p}}(\Delta_{\bm{p}} \pm \Delta')$.  The 
Hamiltonian in the new basis is
\begin{equation}
H_{0} = \sum_{\bm{p}} \left(\varepsilon_{\bm{p}} c^{\dag}_{\bm{p}}c_{\bm{p}} +\Delta_{\bm{p}} c_{\bm{p}}^{\dag}\sigma_{z}c_{\bm{p}}\right),
\end{equation}
which has two energy bands split by the energy gap $\Delta_{\bm{p}} =
\sqrt{l_{\bm{p}}^{2} - 2l_{\bm{p}}^{z}\Delta + \Delta^{2}}$ due the exchange and
spin-orbit couplings.

Interaction of the conduction band with the dynamic magnetization in Eq.~\eqref{eq:H0} is given by
\begin{equation}
H_{i} = -\Delta \sum_{\bm{p}} \left( \delta S_{x} \psi^{\dag}_{\bm{p}} \sigma_{x} \psi_{\bm{p}} + \delta S_{y} \psi^{\dag}_{\bm{p}} \sigma_{y} \psi_{\bm{p}} \right).
\end{equation}
We use a linearized Holstein-Primakoff representation \cite{Holstein1940} for
the dynamic spin components, $\delta S_{x} = \sqrt{1/2}(a + a^{\dag})$ and
$\delta S_{y} = i\sqrt{1/2}(a - a^{\dag})$, where $a$ and $a^{\dag}$  satisfy
boson commutation rules.  After the unitary transformation in Eq.~\eqref{eq:Up},
the interaction part of the Hamiltonian becomes
\begin{equation}
	H_{i} = \sum_{\bm{p}}\left( \bm{\lambda}_{\bm{p}}\cdot a c^{\dag}_{\bm{p}} \bm{\sigma} c_{\bm{p}} + \bm{\lambda}_{\bm{p}}^{*}\cdot c^{\dag}_{\bm{p}} \bm{\sigma} c_{\bm{p}} a^{\dag} \right),
\end{equation}
where the complex interaction parameter in the transformed frame is given by
\begin{equation}
\bm{\lambda_{p}} = \frac{\Delta}{\sqrt{2}}\frac{l_{\bm{p}}^{(-)}}{l_{\bm{p}}^{\perp}}\left(\frac{\Delta'}{\Delta_{\bm{p}}}, i, -\frac{l_{\bm{p}}^{\perp}}{\Delta_{\bm{p}}}\right),
\label{eq:lamb}
\end{equation}
where $\bm{l}_{\bm{p}}^{\perp} = (l^{x}_{\bm{p}}, l^{y}_{\bm{p}}, 0)$ and $l_{\bm{p}}^{\perp} = |\bm{l}_{\bm{p}}^{\perp}|$.

Lastly, we include the Hamiltonian for the magnetization  precessing with the
the Kittel mode frequency $\Omega$ in a uniform magnetic field applied along the
$z$ axis, $H_{m} = -\hbar\Omega S_{z}$.  Using the representation $S_{z} =
1-a^{\dag}a$ \cite{Holstein1940}, the complete Hamiltonian is written in the
form of a two-band model interacting with an isolated bosonic mode
\begin{multline}
  H = \hbar\Omega a^{\dag}a + \sum_{\bm{p}} \left(\varepsilon_{\bm{p}} c^{\dag}_{\bm{p}}c_{\bm{p}} +\Delta_{\bm{p}} c_{\bm{p}}^{\dag}\sigma_{z}c_{p}\right) \\+ \sum_{\bm{p}} \left(ac_{\bm{p}}^{\dag}(\bm{\lambda}_{\bm{p}}\cdot \bm{\sigma})c_{\bm{p}} + c^{\dag}_{\bm{p}}(\bm{\lambda}_{\bm{p}}^{*}\cdot \bm{\sigma})c_{\bm{p}}a^{\dag}\right),
  \label{eq:H}
\end{multline}
where $c_{\bm{p}} = (c_{\bm{p}\uparrow}, c_{\bm{p}\downarrow})^{T}$, 
$c^{\dag}_{\bm{p}} = (c^{\dag}_{\bm{p}\uparrow}, c^{\dag}_{\bm{p}\downarrow})$,
describe destruction and creation of electrons in the upper ($\uparrow$) or
lower ($\downarrow$) band.  Notice that $\lambda^{z}_{\bm{p}}$ term commutes with $H_{0}$ and couples the boson mode to the difference in band occupation numbers.

\section{Susceptibility for a driven spin-orbit torque oscillator}
\label{sec:tls}
The nature of the dynamic regime of an interacting magnetic oscillator can be
examined by looking at the magnetic susceptibility. Here, we derive the
susceptibility $\chi(\omega)$ for the oscillator in Eq.~\eqref{eq:H} in the
presence of spin-orbit torques produced by the conduction electrons that are
driven externally by applying the field.

The susceptibility $\chi(\omega)$ is defined as the Fourier transform of a
retarded Green's function \cite{Tiablikov2013}
\begin{equation}
\chi(\omega) = -i \int_{0}^{\infty} dt e^{i \omega t - \epsilon t}
\langle [ S^{(-)}(t), S^{(+)}(0) ] \rangle,
\label{eq:chi2}
\end{equation}
where the circular components of the spin operator are $S^{(\pm)}=S_{x} \pm
iS_{y}$.  Here $\langle \ldots \rangle$ denotes averaging with respect to the
density matrix $\rho = \exp(-H/k_{B}T) / \Tr \exp(-H/k_{B}T)$ of the system in
equilibrium, and $\epsilon \to 0^{+}$.

In terms of the model in Eq.~\eqref{eq:H}, this includes calculating the
retarded Green's function for the boson mode
\begin{equation}
	\llangle a(t), a^{\dag}(t') \rrangle = -i\theta(t-t')\Tr\left\lbrace [a(t), a^{\dag}(t')] \rho \right\rbrace,
\end{equation}
where the operators are in the Heisenberg picture $a(t) =
\exp(iHt/\hbar)a\exp(-iHt/\hbar)$.

We can simplify the model to see clearly the physics essential to our
discussion.  We drop the momentum index in Eq.~\eqref{eq:H} and consider a
two-level system interacting with a quantum oscillator
\begin{equation}
H = \Omega a^{\dag}a + \frac{\Delta}{2} s^{z} + a(\bm{\lambda} \cdot \bm{s}) + (\bm{\lambda}^{*}\bm{s})a^{\dag},
\end{equation}
where $\bm{\lambda}$ is a complex interaction parameter, and the electron spin-$\frac{1}{2}$ operators are $s^{\alpha} =
c^{\dag}\sigma_{\alpha}c$. We use $\hbar = 1$ for convenience and restore it
when necessary.

The equation of motion \cite{Tiablikov2013} for the bosonic Green's function is
\begin{multline}  \label{eq:eom2}
i\frac{\partial}{\partial t} \llangle a(t), a^{\dag}(t') \rrangle = \delta(t - t') + \Omega \llangle a(t), a^{\dag}(t') \rrangle \\
+ \bm{\lambda}^{*} \cdot \llangle \bm{s}(t),a^{\dag}(t') \rrangle,
\end{multline}
and is coupled with the equation of motion for the mixed Green's function:
\begin{multline}
  i\frac{\partial}{\partial t} \llangle \bm{s}(t), a^{\dag}(t') \rrangle =
  i\Delta \llangle \left[\hat{\bm{z}} \times  \bm{s}(t)\right], a^{\dag}(t') \rrangle \\
  + 2i \llangle a(t)[\bm{\lambda} \times \bm{s}(t)],a^{\dag}(t') \rrangle \\
  + 2i \llangle \left[\bm{\lambda}^{*} \times \bm{s}(t) \right] a^{\dag}(t),a^{\dag}(t') \rrangle,
  \label{eq:eom}
\end{multline}
where $\hat{\bm{z}}$ is the unit vector along the $z$ axis.

Decoupling is achieved using the approximation, $\llangle a(t) s^{\gamma}(t),
a^{\dag}(t')  \rrangle \approx \langle s^{\gamma}(t) \rangle \llangle a(t),
a^{\dag}(t') \rrangle$, which treats $\langle s^{\gamma}(t) \rangle$ as an
external driving force. This assumption is reasonable if relaxation time of
magnetic dynamics is much larger that the electron scattering time. Similar
situation occurs for  current generated spin transfer torques in magnetic
textures \cite{Kishine2010}.

Using the shorthand notation $G(t,t') = \llangle a(t), a^{\dag}(t') \rrangle$,
$\tilde{G}(t,t') = \llangle a^{\dag}(t), a^{\dag}(t') \rrangle$ and
$\bm{F}(t,t') = \llangle \bm{s}(t), a^{\dag}(t')\rrangle$, we rewrite
Eqs.~\eqref{eq:eom} as
\begin{subequations}
\label{eq:eom1}
\begin{eqnarray}
  i \frac{\partial F_{x}}{\partial t} &=& -i\Delta F_{y} + \mathfrak{T}_{x}G + \mathfrak{T}^{*}_{x}\tilde{G}, \label{eq:Fx} \\
  i \frac{\partial F_{y}}{\partial t} &=&  i\Delta F_{x} + \mathfrak{T}_{y}G + \mathfrak{T}^{*}_{y}\tilde{G}, \label{eq:Fy} \\
  i \frac{\partial F_{z}}{\partial t} &=&                  \mathfrak{T}_{z}G + \mathfrak{T}^{*}_{z}\tilde{G}, \label{eq:Fz}
\end{eqnarray}
\end{subequations}
where we introduce the torque, $\bm{\mathfrak{T}} = 2i [\bm{\lambda} \times
\langle \bm{s}(t) \rangle]$.  This torque is exerted by the quantum oscillator
on the electron system.  This system of equations should be supplemented by the
equation of motion for $\tilde{G}(t,t')$.  However, the contribution from
$\tilde{G}(t,t')$ can be neglected, since we are mostly interested in the region
close to the resonance $\omega \approx \Omega$ (see Appendix~\ref{sec:tG} for
details).

In what follows, we consider a stationary regime where the spin accumulations
and, therefore, the torque $\mathfrak{T}$ are independent of time. In this
situation, $G(t,t')$ and $\bm{F}(t,t')$ depend on $t-t'$ and can be Fourier
transformed according to $G(\omega) = \int_{-\infty}^{\infty} dt \exp(i \omega
t) G(t)$.  The equations of motion \eqref{eq:eom} and \eqref{eq:eom1} in
the frequency domain give
\begin{equation}
  G(\omega) = \frac{1}{\omega - \Omega - \Sigma(\omega)}.
  \label{eq:GF}
\end{equation}
where 
\begin{equation}
 \Sigma(\omega)  =  \frac{1}{2} \frac{\lambda_{(+)}^{*}\mathfrak{T}_{(-)}}{\omega - \Delta} + \frac{1}{2} \frac{\lambda_{(-)}^{*}\mathfrak{T}_{(+)}}{\omega + \Delta} + \frac{\lambda_{z}^{*}\mathfrak{T}_{z}}{\omega}, \label{eq:SE} 
\end{equation}
with $\lambda_{(\pm)} = \lambda_{x} \pm i\lambda_{y}$ and $\mathfrak{T}_{(\pm)}
=  \mathfrak{T}_{x} \pm i\mathfrak{T}_{y}$.

The self-energy in Eq.~\eqref{eq:SE} has two different contributions. The first
contribution exists for thermodynamic equilibrium where $\langle s^{x} \rangle =
\langle s^{y} \rangle = 0$. Equilibrium is given by
\begin{equation}  \label{eq:Seq}
\Sigma_{\mathrm{eq}}(\omega) = \frac{2\langle s^{z} \rangle}{\omega^{2} - \Delta^{2}} \left[i\omega(\bm{\lambda}^{*} \times \bm{\lambda})_{z} - \Delta(|\lambda_{x}|^{2} + |\lambda_{y}|^{2})\right],
\end{equation}
where $\langle s^{z} \rangle = f_{\uparrow} - f_{\downarrow}$ with $f_{\sigma}$
being the population of the $\sigma = (\uparrow, \downarrow)$ energy level.

Close to the resonance $\omega \approx \Delta$, we can approximate $\omega + \Delta \approx 2\Delta$, and Eq.~(\ref{eq:Seq}) can be written in a simplified form
\begin{equation}
\label{eq:Keq}
\Sigma_{\mathrm{eq}}(\omega) = \frac{K_{\mathrm{eq}}}{\omega - \Delta}, 
\end{equation}
where $K_{\mathrm{eq}} = \langle s^{z} \rangle \left[ i(\lambda_{x}^{*}
\lambda_{y} - \lambda_{y}^{*}\lambda_{x}) - (|\lambda_{x}|^{2} +
|\lambda_{y}|^{2})\right]$ is manifestly \emph{positive} because in equilibrium
$f_{\uparrow} < f_{\downarrow}$.

Out of thermodynamic equilibrium, a transverse spin accumulations  $\langle  s^{x} \rangle$ and $\langle s^{y} \rangle$ can appear.  The self-energy then acquires an additional contribution $\Sigma(\omega) = \Sigma_{\mathrm{eq}}(\omega) + \delta\Sigma(\omega)$ from the torques $\mathfrak{T}_{(\pm)}$.   
In this decoupling scheme, $\delta\Sigma(\omega)$ is the second order term, which mixes $\lambda_{z}$ with interband spin transitions.  In the region $\omega \approx \Delta$, this additional contribution can be approximated as
\begin{equation}
\label{eq:Kne}
\delta\Sigma(\omega) = \frac{\lambda_{z}\lambda_{(+)}^{*}\langle\delta s^{(-)}\rangle}{\omega - \Delta},
\end{equation}
where $\langle\delta s^{(-)}\rangle = \langle s^{x} \rangle - i \langle s^{y}
\rangle$.  We note that in contrast to $\delta\Sigma$, interband interactions $\lambda_{x}$ and $\lambda_{y}$ fully determine coupling to the oscillator mode in $\Sigma_{\mathrm{eq}}$.

According to Eqs.~\eqref{eq:Keq} and \eqref{eq:Kne}, the magnetic susceptibility
that describes interaction between resonances in the electron and magnetic
subsystems is written in the following form
\begin{equation} \label{eq:chi0}
 \chi(\omega) = 2\left[\omega - \Omega + i\Gamma - \frac{K}{\omega - \Delta 
 	+i \gamma}\right]^{-1},
\end{equation}
where $K = K_{\mathrm{eq}} + \delta K$, and  $\delta K$ is determined by Eq.~(\ref{eq:Kne}). Here, we included $\Gamma$ and $\gamma$, which now have a meaning of phenomenological relaxation parameters for the magnetic and electron channels respectively.

Equations (\ref{eq:Keq}) and (\ref{eq:Kne}) show that in an open system the parameter $K$ that
describes interaction between two resonances is generally complex as a result of
external driving.  Interaction of two modes near the energy level crossing in this situation
can be described by a linear non-Hermitian Hamiltonian in Eq.~(\ref{eq:osc})
with a complex coupling $ g = \sqrt{K} $.

\section{Mode attraction in a spin-orbit torque system driven by electric field}
\label{sec:lat}
We apply our general results from Sec.~\ref{sec:tls} to the model of the spin-orbit torque system in Eq.~\ref{eq:H}. Using Eqs.~(\ref{eq:Keq}) and (\ref{eq:Kne}), and restoring the momentum index, we find the following expression for the self-energy of the magnetic oscillator
\begin{multline}
  \Sigma(\omega) =  -\frac{1}{2}\sum_{\bm{p}} \frac{\Delta^{2}}{\omega - 2\Delta_{\bm{p}}}\left[ \left(1 + \frac{\Delta'}{\Delta_{\bm{p}}}\right)^{2}\langle s^{z}_{\bm{p}} \rangle \right. \\
  \left.
  + \frac{l_{\bm{p}}^{\perp}}{\Delta_{\bm{p}}} \left(1 + \frac{\Delta'}{\Delta_{\bm{p}}}\right)\langle \delta s^{(-)}_{\bm{p}} \rangle \right],
  \label{eq:K}
\end{multline}
where $\langle \delta s^{\alpha}_{\bm{p}} \rangle = \langle c_{\bm{p}}^{\dag} \sigma_{\alpha} c_{\bm{p}} \rangle$ for $\alpha = x, y$ can be interpreted as an average rate of interband electron transitions between  $\varepsilon_{\bm{p}\sigma} = \varepsilon_{\bm{p}} \pm \Delta_{\bm{p}}$ bands, where upper (lower) sign is for $\sigma = \uparrow$ ($\sigma = \downarrow$), while $\langle s_{\bm{p}}^{z} \rangle =  f_{\bm{p}\uparrow} - f_{\bm{p}\downarrow}$ denotes the difference in band occupation numbers with $f_{\bm{p}\sigma} = \lbrace \exp \left[ (\varepsilon_{\bm{p}\sigma} - \mu)/k_{B}T \right] +1 \rbrace^{-1}$ being the Fermi-Dirac distribution with the temperature $T$ and chemical potential $\mu$.

We note that the self-energy in Eq.~(\ref{eq:K}) has a form of the self-energy of the Fano-Anderson model \cite{Mahan2013}, $\Sigma(\omega) = \sum_{\bm{p}} g^{2}_{\bm{p}}/(\omega - 2\Delta_{\bm{p}})$ with complex coupling $g_{\bm{p}}$. General analysis of this model is beyond the scope of this paper and will be studied elsewhere.  For the rest of discussion, we use the strong exchange interaction approximation, $\Delta \gg l_{\bm{p}}$  \cite{Manchon2008}.  In this regime, the energy splitting between two bands is independent of the momentum, $\Delta_{\bm{p}}\approx \Delta$, so that the susceptibility of the spin-orbit torque oscillator is given by Eq.~\eqref{eq:chi0} with
\begin{equation}
	\label{eq:K1}
  K = -2\Delta^{2}\sum_{\bm{p}} \left( \langle s^{z}_{\bm{p}}\rangle + \frac{l_{\bm{p}}}{2\Delta} \langle \delta s^{(-)}_{\bm{p}} \rangle  \right).
\end{equation}
In this equation, $\langle \delta s^{(-)}_{\bm{p}} \rangle = \langle \delta s^{x}_{\bm{p}} \rangle - i\langle \delta s^{y}_{\bm{p}} \rangle$ can be interpreted as the transverse electron spin accumulation in the local frame 
\footnote{Neglecting terms of order $l_{\bm{p}}/\Delta$ means that $U_{\bm{p}}^{\dag} \sigma_{x} U_{\bm{p}} = -\cos \varphi_{\bm{p}} \sigma_{x} - \sin \varphi_{\bm{p}} \sigma_{y}$ and $U_{\bm{p}}^{\dag} \sigma_{y} U_{\bm{p}} = -\sin \varphi_{\bm{p}} \sigma_{x} + \cos \varphi_{\bm{p}} \sigma_{y}$, where $\tan \varphi_{\bm{p}} = l^{y}_{\bm{p}}/l^{x}_{\bm{p}}$, see Eqs.~(\ref{eq:sx})--(\ref{eq:sz}).  These equations show that $x$ and $y$ components of the electron spin in the laboratory frame are related to the spin accumulations in Eq.~(\ref{eq:K}) by the $SU(2)$ rotation.}.

Spin accumulations can be induced, for example, via the inverse spin galvanic effect \cite{Dyakonov1971, Edelstein1990}.
Since spin-orbit interaction mixes spin and orbital degrees of freedom, a static electric field is able to induce spin accumulations in systems with broken inversion symmetry that, in turn, can exert a torque on a magneic oscillator \cite{Manchon2019}.

We estimate these spin accumulations using a linear response approach in Appendix~\ref{app:B}. In the limit where the band splitting $\Delta$ is much greater than broadening due to impurity scattering, these spin accumulations are given by
\begin{subequations}
\label{eq:st}
\begin{eqnarray}
	\label{eq:st1}
	\langle \delta s_{\bm{p}}^{x} \rangle &=& 
	\frac{e(f_{\bm{p}\uparrow} - f_{\bm{p}\downarrow})}{2\Delta^{2}l_{\bm{p}}}  
	\left[\bm{l}_{\bm{p}} \times \frac{\partial \bm{l}_{\bm{p}}}{\partial p_{\alpha}}\right]_{z} E_{\alpha}, \\ 
	\langle \delta s_{\bm{p}}^{y} \rangle &=& \frac{e(f_{\bm{p}\uparrow} - f_{\bm{p}\downarrow})}{2\Delta^{2}l_{\bm{p}}} \left(\bm{l}_{\bm{p}} \cdot \frac{\partial \bm{l}_{\bm{p}}}{\partial p_{\alpha}}\right) E_{\alpha},
\end{eqnarray}
\end{subequations}
where $\bm{E}$ is the electric field strength, and $e$ is the electron charge.  In the transformed frame, these quantities play different roles: $\langle \delta s_{\bm{p}}^{y} \rangle$ contributes $\Im K$ and is involved mostly into the energy dissipation balance, see Eq.~(\ref{eq:ex2}), while $\langle \delta s_{\bm{p}}^{x} \rangle$ can be interpreted as a renormalization of the coupling constant between two resonances.  Equations (\ref{eq:st1}) show that to create $\langle \delta s_{\bm{p}}^{x} \rangle$, the spin-orbit field $\bm{l}_{\bm{p}}$ should have at least two components.

Changing the sign of $\Re K$ drives the system toward instability. Since spin accumulations are proportional to $f_{\bm{p}\uparrow} - f_{\bm{p}\downarrow}$, the main contribution in Eq.~(\ref{eq:K1}) comes from the Fermi surface.  To reach the instability regime, the following condition should be satisfied on the Fermi surface
\begin{equation}
    \label{eq:cr}
	1 + \frac{eE_{\alpha}}{\Delta^{3}} \left[\bm{l}_{\bm{p}} \times
	\frac{\partial \bm{l}_{\bm{p}}}{\partial p_{\alpha}}\right]_{z} <0. 
\end{equation}
The $z$ component of the axial vector $\bm{l}_{\bm{p}} \times \partial (\bm{l}_{\bm{p}}/\partial p_{\alpha}) E_{\alpha}$ in this expression can be interpreted as an effective field acting on the electron spin that competes with the exchange field $\Delta$ \footnote{In other words, the electric-field-induced interband matrix elements of the electron current should be compared to the band gap $\Delta$}.

In order to reach the mode attraction regime, we need special symmetry conditions. First, since the second term in Eq.~(\ref{eq:cr}) is odd in $\bm{p}$, we need the original band to be noncentrosymmetric, $\varepsilon_{\bm{p}} \ne \varepsilon_{-\bm{p}}$, in order to preserve the contributions from Eqs.~(\ref{eq:st}) after the summation over the Brillouin zone.  Second, when $\langle \delta s_{\bm{p}}^{y} \rangle \ll \langle \delta s_{\bm{p}}^{x} \rangle$, level attraction can occur, as shown in Fig.~\ref{fig:1}~(b). We show how this can occur for a spin-orbit torque oscillator with Rashba coupling in the next section.

\begin{figure}[t]
	\centerline{\includegraphics[width=0.49\textwidth]{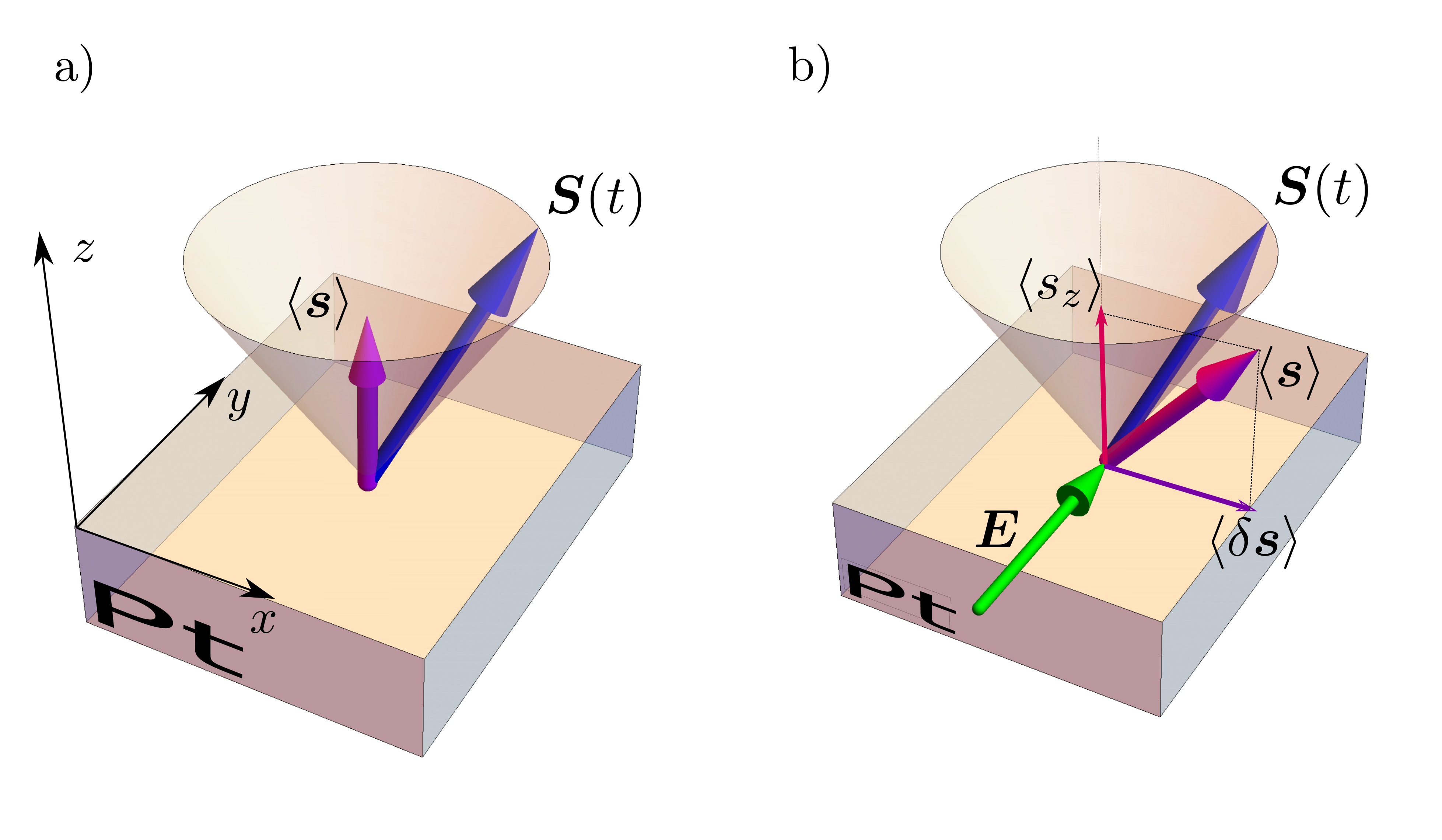}}
	\caption{Schematic picture of a magnetic oscillator $\bm{S}(t)$ coupled to the material (Pt) with Rashba spin-orbit interaction $\frac{\alpha}{\hbar}(\bm{p} \times \hat{\bm{z}})\cdot \bm{\sigma}$ via $s$-$d$ exchange coupling. (a) In equilibrium, $\bm{S}(t)$ precesses around the quantization $z$ axis, which coincides with the direction of the magnetization of conduction electrons $\langle \bm{s} \rangle$. (b) Applying electric field $\bm{E}$ in noncentrosymmetric material creates a transverse spin accumulation $\langle \delta \bm{s} \rangle$ that modifies dynamics of the magnetic oscillator.}
	\label{fig:2}
\end{figure}

\section{Realization for a spin-orbit torque oscillator with Rashba coupling}
\label{sec:res}
A physical system where our approach can be used is the spin-orbit torque oscillator with Rashba type of spin-orbit interaction.  In two spatial dimensions, the Rashba model is described by the Hamiltonian in Eq.~\eqref{eq:H0} with 
\begin{equation}
\bm{l}_{\bm{p}} = \frac{\alpha}{\hbar} (\bm{p} \times \hat{\bm{z}}),
\end{equation}
where $\bm{p} = (p_{x}, p_{y}, 0)$ and $\alpha$ is the Rashba spin-orbit coupling \cite{Bychkov1984}. Schematically, this setup is illustrated in Fig.~\ref{fig:2}.  The spin accumulations in Eqs.~\eqref{eq:st} are reduced to 
\begin{subequations}
	\label{eq:sr}
\begin{eqnarray}
  \langle \delta s_{\bm{p}}^{x} \rangle &=&\frac{e\alpha}{\hbar} \left(f_{\bm{p}\uparrow} - f_{\bm{p}\downarrow}\right) \frac{\bm{E}\cdot(\hat{\bm{z}}\times\bm{p})}{2 p\Delta^{2}}, \\
  \langle \delta s_{\bm{p}}^{y} \rangle &=& \frac{e\alpha}{\hbar} \left(f_{\bm{p}\uparrow} - f_{\bm{p}\downarrow}\right) \frac{(\bm{E} \cdot \bm{p})}{2 p\Delta^{2}}.
\end{eqnarray}
\end{subequations}
We note that for the Rashba model $\langle \delta s_{\bm{p}}^{x} \rangle$ is proportional to $\alpha (\bm{p} \times \bm{E})_{z}$, which may be interpreted as the $z$ component of the relativistic magnetic field in the coordinate frame of the moving electrons. Interestingly, when this field is strong enough to satisfy Eq.~(\ref{eq:cr}), system behavior becomes similar the two-oscillator model with negative-energy modes \cite{Bernier2018}.

The electric field in Eqs.~(\ref{eq:sr}) couples to different components of the momentum. For example, if we apply the field along the $y$ axis, $\langle \delta s_{\bm{p}}^{y} \rangle$ becomes proportional to $p_{y}$, while $\langle \delta s_{\bm{p}}^{x} \rangle$ is determined by $p_{x}$.  This can be used to manipulate relative contributions to real and imaginary parts of $K$ by rotating the system along the $z$ axis.

Suppose that we deal with a noncentrosymmetric band $\varepsilon_{\bm{p}} \ne \varepsilon_{-\bm{p}}$, in a situation where there is a mirror symmetry with respect to $p_{y} \to -p_{y}$. In this case, the contribution from $p_{y}$ vanishes, and the effective interaction between two resonance in Eq.~(\ref{eq:K1}) is estimated as follows 
\begin{equation} \label{eq:K12}
  K = -\Delta^{3} \sum_{\bm{p}}\left(f_{\bm{p}\uparrow} - f_{\bm{p}\downarrow}\right)\left(1 - \frac{e \alpha^{2}E p_{x} \exp(i\beta)}{\hbar\Delta^{3}}\right),
\end{equation}
where $\beta$ denote the angle between the electric field and $y$ axis. In a situation where $\Delta$ is much less than the Fermi energy $\varepsilon_{F}$, we can estimate $\Re K \approx g(\varepsilon_{F})S_{0}^{-1}(\Delta^{3} - e \xi \alpha^{2}p_{F}E\hbar^{-1})$, where $g(\varepsilon_{F})$ is the density of states at the Fermi level, $p_{F}$ denotes the Fermi momentum, and $\xi \lesssim 1$ and the dimensionless geometric factor that comes from the integration over the Fermi surface and reflect noncentrosymmetry of the band $\varepsilon_{\bm{p}}$.  To reach the attractive regime, we need the electric field strength to be above the critical value, which makes $\Re K = 0$ at the threshold. This field is estimated as
\begin{equation}
	E_{c} =  \frac{\Delta^{3}}{e\alpha^{2}k_{F}},
\end{equation}
where $k_{F}$ is the Fermi momentum.

\begin{figure}[t]
	\centerline{\includegraphics[width=0.35\textwidth]{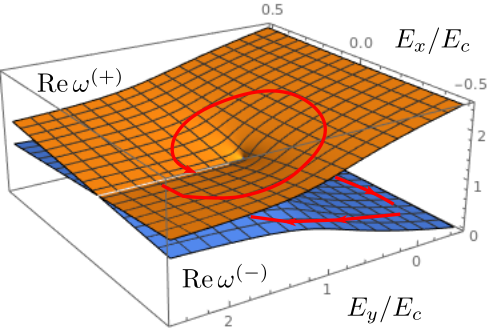}}
	\caption{Real parts of $\omega^{(+)}$ and $\omega^{(-)}$ for the spin-orbit torque oscillator with the Rashba spin-orbit interaction on the $(E_{x}, E_{y})$ plane. The branch cut, which is determined by Eq.~(\ref{eq:ex2}), terminates at the exceptional point given by Eq.~(\ref{eq:ex1}). A closed loop around the exceptional point is shown by the red line.}
	\label{fig:3}
\end{figure}

For the electric field applied along the $y$-direction ($ \beta = 0 $), the contribution from $\langle s_{\bm{p}}^{y} \rangle$ vanishes, and $K$ is purely real. In contrast, applying the electric field along the mirror plane ($ \beta = \pi/2 $) will remove the contribution from  $\langle \delta s_{\bm{p}}^{x} \rangle$. Therefore, by changing the direction of the electric field and its strength, we can in principle satisfy both Eqs.~(\ref{eq:ex}) and thus reach the exceptional point. We plot the real parts of $\omega^{(\pm)}$ on the $(E_{x}, E_{y})$ plane in Fig.~\ref{fig:3}. This surface has a branch cut determined by Eq.~(\ref{eq:ex2}). By manipulating $E_{x}$ and $E_{y}$ it is possible to trace a closed loop around the exceptional point. For example, starting from the lower branch and then going through the branch cut, we would have $\omega^{(-)} \to \omega^{(+)} \to \omega^{(+)} \to \omega^{(-)}$ \cite{Heiss2004, Heiss2012}, as shown in Fig.~\ref{fig:3}.

This opens a possible way for an electric-field switching of the oscillator regime. Suppose that at $E = 0$ we have two hybridized modes with the frequencies $\omega^{(\pm)} \approx \Omega \pm \sqrt{(\delta/2)^{2} + K_{\mathrm{eq}} }$ where $\delta = \Omega - \Delta$ is the mode detuning parameter, and $K_{\mathrm{eq}}$ is a positive mode hybridization.  Starting for example with the lower mode $\omega^{(-)}$ and slowly varying the electric field above $E_{c}$, we can go trough the branch cut (see Fig.~\ref{fig:3}), and after that decrease the electric field back to $E = 0$. As a result, the systems will oscillate with the frequency $\omega^{(+)}$, which is normally separated from our initial mode $\omega^{(-)}$ by the energy gap of order $\sqrt{K_{\mathrm{eq}}}$.

For experimental observation, we need $\Delta$ withing the same range as the frequency of the ferromagnetic resonance $\Omega$. In typical spin-orbit torque materials \cite{Manchon2008}, $\alpha \approx 10^{-12}$~eV/m, and $k_{F} \approx 10^{9}$~m$^{-1}$, so that $\alpha k_{F}/\Delta \lesssim 1$. This allows to estimate the minimal electric field strength as $\Delta k_{F}/e \approx 10^{4}$~V/cm.

\begin{figure}[t]
	\centerline{\includegraphics[width=0.485\textwidth]{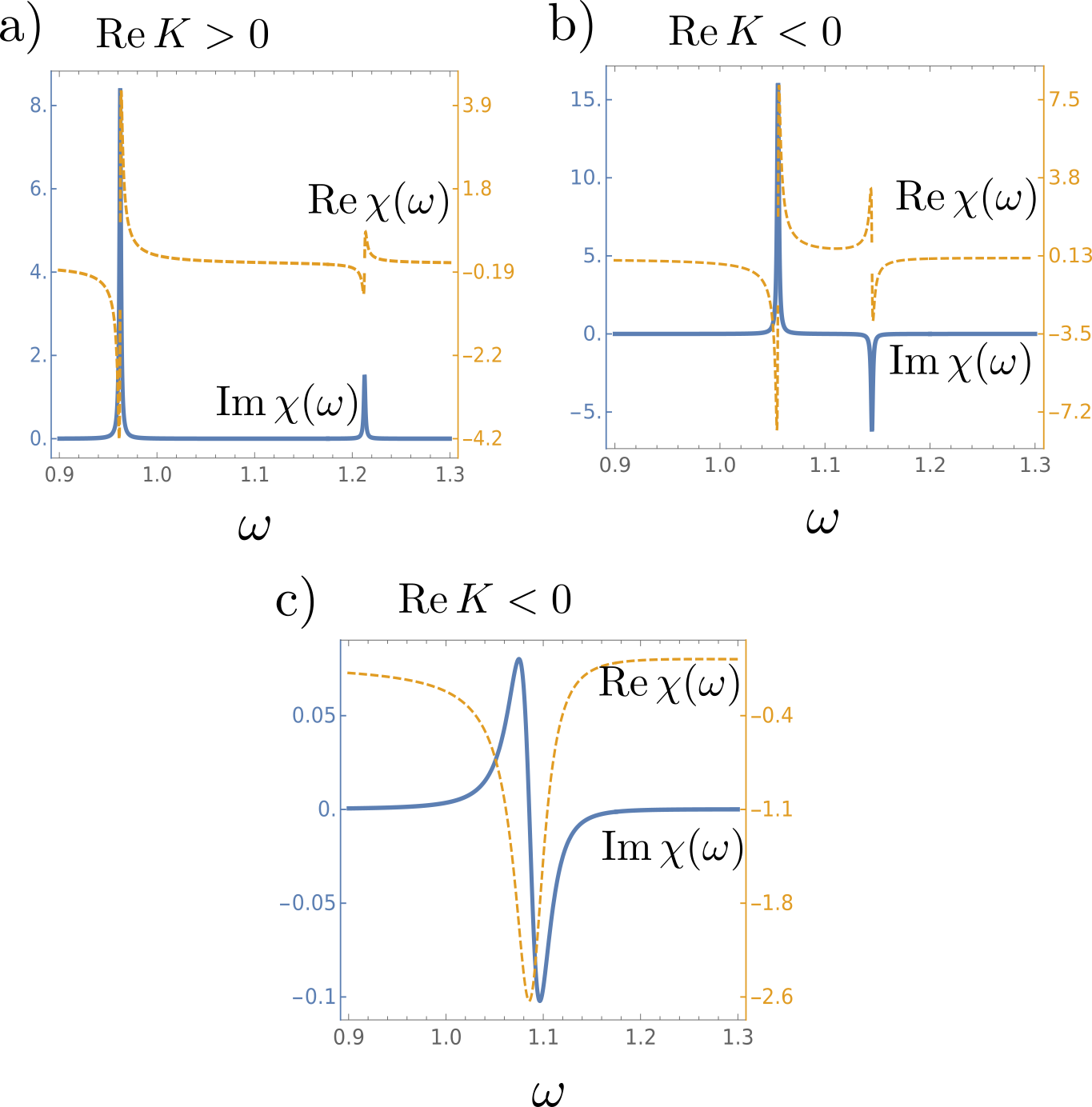}}
	\caption{Magnetic susceptibility for the Rashba spin orbit torque oscillator (arbitrary units). Solid (dashed) line shows the imaginary (real) part of $\chi(\omega)$. (a) Two absorption peaks for $E < E_{c}$. (b) Resonance and anti-resonance in the case of $E > E_{c}$ and weak interaction $|\Omega - \Delta| \gtrsim |\Re K|$. (c) Strongly interacting case for $E > E_{c}$ and $|\Omega - \Delta| \lesssim |\Re K|$.}
	\label{fig:4}
\end{figure}

Experimentally, mode attraction can be identified by measuring susceptibility of the magnetic oscillator in the electric field.  Below the critical field, one has usual behavior of $\Im \chi(\omega)$ showing two absorption peaks, as schematically demonstrated in Fig.~\ref{fig:4}~(a).  In contrast, when $E > E_{c}$, the second resonance associated with electron transitions changes its sign, see Fig.~\ref{fig:4}~(b). The behavior of $\Im \chi(\omega)$ becomes particularly interesting in the region $|\Omega - \Delta| \lesssim |\Re K| $ where the two resonances strongly interact. In this situation, $\Im \chi(\omega)$ shows a sign change, while $\Re \chi(\omega)$ has a peak (Fig.~\ref{fig:4}~c). Interestingly, similar behavior of a response coefficient has been observed for a spin orbit torque system in a microwave waveguide \cite{Berger2018}.

\section{Conclusions}
\label{sec:con}

Mode attraction can be observed in a Rashba spin-orbit torque system driven by an electric field.  In noncentrosymmertic materials, the electric field couples to orbital motion of conduction electrons and thus creates an effective magnetic field, which renormalizes the interaction between electron and magnetic subsystems. In the case when the field strength is above the critical value $E_{c}$, the interaction between electron and magnetic resonances near the energy level crossing can be effectively described by a non-Hermitian Hamiltonian with a complex coupling constant. The real and imaginary parts of these constant can be separately controlled by different component of the electric field, when special symmetry conditions are met. This allows to realize an exceptional point in the energy spectrum of a driven dissipative spin-orbit torque system, and explore its topological structure by manipulating the electric field.  This opens a way for electric-field selection and switching of a magnetic oscillator frequency.

Similar behavior may be expected in spin-transfer torque systems. However, detailed analysis of this topic is beyond the scope of this paper and requires a separate investigation.

\acknowledgements{
IP and RLS acknowledge the support of the Natural Sciences and Engineering
Research Council of Canada (NSERC) RGPIN 05011-18.}

\appendix

\section{Equation of motion for $\tilde{G}(t,t')$}
\label{sec:tG}
Equations~\eqref{eq:eom2} and \eqref{eq:eom1} should be supplemented by the
equation of motion for $\tilde{G}(t,t')$:
\begin{equation}
i \frac{\partial \tilde{G}}{\partial t} = -\Omega \tilde{G} - \bm{\lambda} \cdot \bm{F}.
\end{equation}
Solving this equation together with Eqs~\eqref{eq:eom2} and \eqref{eq:eom1} in
the $\omega$-domain gives
\begin{eqnarray}
  \left[\omega - \Omega - \Sigma(\omega)\right]G(\omega) - A^{*}(\omega) \tilde{G}(\omega) &=& 1,\\
  \left[\omega + \Omega + \Sigma^{*}(\omega)\right]\tilde{G}(\omega) + A(\omega)G(\omega)   &=& 0,
\end{eqnarray}
where we also introduce
\begin{equation}
 A(\omega) = \frac{1}{2} \frac{\lambda_{(+)}\mathfrak{T}_{(-)}}{\omega - \Delta} + \frac{1}{2} \frac{\lambda_{(-)}\mathfrak{T}_{(+)}}{\omega + \Delta} + \frac{\lambda_{z}\mathfrak{T}_{z}}{\omega}.
\end{equation}
The solution for $G(\omega)$ is as follows
\begin{equation}
G(\omega) = \frac{\omega + \Omega + \Sigma^{*}(\omega)}{(\omega - \Omega - \Sigma(\omega))(\omega + \Omega + \Sigma^{*}(\omega))+|A(\omega)|^{2}},
\end{equation}
which near $\omega \approx \Omega$ can be approximated by Eq.~\eqref{eq:GF}.

\section{Spin accumulations in the linear response regime}
\label{app:B}
We outline calculation of the transverse spin accumulations $\langle
\bm{s}^{\alpha}_{\bm{p}} \rangle$ ($\alpha = x,y$).  For this purpose, we
consider the linear response of the electron system with the Hamiltonian in
Eq.~(\ref{eq:H1}) to the static electric field $\bm{E}$.  In the frame where
$H_{0}$ is diagonal, spin accumulations are given by the following Kubo formula
\cite{Mahan2013}
\begin{equation}
  \langle s_{\bm{p}}^{\alpha} \rangle = 
  \int_{-\infty}^{0} dt e^{st} \int_{0}^{\beta} d\tau 
  \Tr \left[ \rho_{0} J^{\beta}(t - i\tau) s_{\bm{p}}^{\alpha} \right] E_{\beta},
  \label{eq:sa}
\end{equation}
where $\beta = (k_{B}T)^{-1}$, $s \to 0^{+}$, $\bm{J}$ is the electric current
operator in the Heisenberg picture, and $\rho_{0}$ denotes density matrix for
the noninteracting system.

For the Hamiltonian in Eq.~(\ref{eq:H1}), the current operator is given by the
expression
\begin{equation}
J_{\alpha} = \sum_{\bm{p}} \frac{\partial \varepsilon_{\bm{p}}}{\partial p_{\alpha}}  \psi^{\dag}_{\bm{p}} \psi_{\bm{p}} + \sum_{\bm{p}} \frac{\partial \bm{l}_{\bm{p}}}{\partial p_{\alpha}} \cdot \psi^{\dag}_{\bm{p}} \bm{\sigma} \psi_{\bm{p}},
\label{eq:J0}
\end{equation}
which contains interband matrix elements proportional to the spin-orbit
coupling.  In the diagonal frame, the matrix elements of the current can be
found using the following transformation rules for the Pauli matrices
\begin{eqnarray}
  \label{eq:sx}
  U^{\dag}\sigma_{x}U &=& -\frac{l_{\bm{p}}^{x}}{l_{\bm{p}}^{\perp}} \frac{\Delta'}{\Delta_{\bm{p}}} \sigma_{x} - \frac{l_{\bm{p}}^{y}}{l_{\bm{p}}^{\perp}} \sigma_{y} + \frac{l_{\bm{p}}^{x}}{\Delta_{\bm{p}}}\sigma_{z}, \\
  U^{\dag}\sigma_{y}U &=& -\frac{l_{\bm{p}}^{y}}{l_{\bm{p}}^{\perp}} \frac{\Delta'}{\Delta_{\bm{p}}} \sigma_{x} + \frac{l_{\bm{p}}^{x}}{l_{\bm{p}}^{\perp}} \sigma_{y} + \frac{l_{\bm{p}}^{y}}{\Delta_{\bm{p}}}\sigma_{z},\\ \label{eq:sz}
  U^{\dag}\sigma_{z}U &=& -\frac{l_{\bm{p}}^{\perp}}{\Delta_{\bm{p}}} \sigma_{x} - \frac{\Delta'}{\Delta_{\bm{p}}}\sigma_{z}.
\end{eqnarray}

In the transformed frame, only off-diagonal matrix elements of the current
operator, $J_{\alpha} = \sum_{\bm{p}}c^{\dag}_{\bm{p}\sigma}
(\mathcal{J}^{\alpha}_{\bm{p}})_{\sigma\sigma'} c_{\bm{p}\sigma'}$, contribute
to the transverse spin accumulations in Eq.~(\ref{eq:sa}).  With the help of
Eqs.~(\ref{eq:sx})--(\ref{eq:sz}), the off-diagonal matrix elements of
the current are given by
\begin{multline}
\label{eq:Jt}
  \mathcal{J}_{\bm{p}}^{\alpha} = -\frac{\Delta'}{\Delta_{\bm{p}}}
  \frac{1}{l^{\perp}_{\bm{p}}} \left(\bm{l}^{\perp}_{\bm{p}} \cdot \frac{\partial \bm{l}^{\perp}_{\bm{p}}}{\partial p_{\alpha}} \right) \sigma_{x}
 \\
+\frac{1}{l^{\perp}_{\bm{p}}} \left(\bm{l}^{\perp}_{\bm{p}} \times \frac{\partial \bm{l}^{\perp}_{\bm{p}}}{\partial p_{\alpha}} \right)_{z} \sigma_{y} 
  -\frac{l^{\perp}_{\bm{p}}}{\Delta_{\bm{p}}}\frac{\partial l^{z}_{\bm{p}}}{\partial p_{\alpha}} \sigma_{x}.
\end{multline}

After little algebra, we find the following expression for the spin accumulations from Eq.~\eqref{eq:sa}
\begin{equation}
  \langle s^{\alpha}_{\bm{p}} \rangle  = iE_{\beta} \sum_{\sigma\sigma'} \frac{f_{\bm{p}\sigma} - f_{\bm{p}\sigma'}}{\varepsilon_{\bm{p}\sigma} - \varepsilon_{\bm{p}\sigma'}} \frac{(\mathcal{J}^{\beta}_{\bm{p}})_{\sigma\sigma'}(\sigma_{\alpha})_{\sigma'\sigma}}{\varepsilon_{\bm{p}\sigma} - \varepsilon_{\bm{p}\sigma'} -is}.
\end{equation}
Taking $s \to 0$ limit, we find the following expressions for the transverse spin accumulations
\begin{eqnarray}
  \langle s_{\bm{p}}^{x} \rangle &=& - \frac{2(f_{\bm{p}\uparrow} - f_{\bm{p}\downarrow})}{(\varepsilon_{\bm{p}\uparrow} - \varepsilon_{\bm{p}\downarrow})^{2}} \Im (\mathcal{J}^{\alpha}_{\bm{p}})_{\uparrow\downarrow} E_{\alpha},\\
  \langle s_{\bm{p}}^{y} \rangle &=& - \frac{2(f_{\bm{p}\uparrow} - f_{\bm{p}\downarrow})}{(\varepsilon_{\bm{p}\uparrow} - \varepsilon_{\bm{p}\downarrow})^{2}} \Re (\mathcal{J}^{\alpha}_{\bm{p}})_{\uparrow\downarrow} E_{\alpha},
\end{eqnarray}
which together with Eq.~(\ref{eq:Jt}) gives Eqs.~(\ref{eq:st}) in the limit $\Delta_{\bm{p}} \approx \Delta$ and $l^{z}_{\bm{p}} = 0$.

\bibliography{resonant}

\end{document}